\documentclass[aps,pre,preprint,onecolumn,citeautoscript,superscriptaddress,
eqsecnum]{revtex4-1}
\usepackage{latexsym}
\usepackage{amsmath}
\usepackage{color}
\usepackage{graphicx}
\usepackage{dcolumn}
\usepackage{bm}
\usepackage{txfonts}
\usepackage{mathrsfs}
\usepackage{feynmf}
\usepackage{comment}
\usepackage{float}

\newcommand{\beq}{\begin{equation}}
\newcommand{\eeq}{\end{equation}}
\def\bea{\begin{eqnarray}}
\def\eea{\end{eqnarray}}
\newcommand{\nn}{\nonumber \\}

\newcommand{\vac}{\vert 0 \rangle}

\def\bea{\begin{eqnarray}}
\def\eea{\end{eqnarray}}

\setcitestyle{square}

\begin{document}


\title{Electronic quasiparticles in the quantum dimer model:\\ density matrix renormalization group results}

\author{Junhyun Lee}

\affiliation{Department of Physics, Harvard University, Cambridge MA 02138, USA}

\author{Subir Sachdev}

\affiliation{Department of Physics, Harvard University, Cambridge MA 02138, USA}
\affiliation{Perimeter Institute for Theoretical Physics, Waterloo, Ontario, Canada N2L 2Y5}

\author{Steven R. White}

\affiliation{Department of Physics and Astronomy, University of California, Irvine, CA 92697-4575, USA}

\date{\today\\
\vspace{0.4in}}

\begin{abstract}%
We study a recently proposed quantum dimer model for the pseudogap metal state of the cuprates. The model contains bosonic dimers, representing a spin-singlet valence
bond between a pair of electrons, and fermionic dimers, representing a quasiparticle with spin-1/2 and charge $+e$.  By density matrix renormalization group calculations on a long but finite cylinder, we obtain the ground-state density distribution of the fermionic dimers for a number of different total densities. From the Friedel oscillations at open boundaries, we deduce that the Fermi surface consists of small hole pockets near $(\pi/2, \pi/2)$, and this feature persists up to a doping density of 1/16. 
We also compute the entanglement entropy and find that it closely matches the sum of the entanglement entropies of a critical boson and a low density of free fermions.
Our results support the existence of a fractionalized Fermi liquid in this model.\\
\\
\end{abstract}

\maketitle

\section{Introduction}

A recent paper \cite{Punk15} has proposed a simple quantum dimer model for the pseudogap metal state of the
hole-doped cuprates. The objective of this model is to describe a metal with electronlike quasiparticles, carrying spin-1/2
and charge $e$ but with a Fermi volume which violates the Luttinger theorem for a Fermi liquid (FL). In particular, doping a density
of $p$ holes away from a half-filled insulator should yield, in Fermi liquid theory, a hole Fermi surface of size $1+p$.
Indeed, just such a Fermi surface is observed at large $p$ \cite{Dama05}. 
However, for $p \approx 0.1$, in the pseudogap metals, many 
physical properties are well described by a model of electronlike quasiparticles with a Fermi surface of size $p$ \cite{Nambu}. 
Such a Fermi surface can be obtained in a ``fractionalized Fermi liquid'' (FL*) \cite{TSSSMV03,TSMVSS04}. The model of Ref.~[\onlinecite{Punk15}]
was designed to yield a FL* state with a Fermi surface of size $p$ using ingredients that are appropriate for a
single-band model of cuprate physics. 

Our paper will present density matrix renormalization group (DMRG) results on the dimer model. 
The exact diagonalization results in Ref.~[\onlinecite{Punk15}] were limited to a lattice size of $8 \times 8$ and a single fermionic dimer. 
Here we study significantly larger systems with up to eight fermions and obtain results on the density distribution
of the fermionic dimers and the entanglement entropy. As we shall see below, all of our results are consistent with 
the appearance of a FL* metal in this dimer model. 

\section{Model and DMRG Setup}

\begin{figure}[t]
\centering
	\vspace*{1mm}
	\includegraphics*[width=60mm,angle=0]{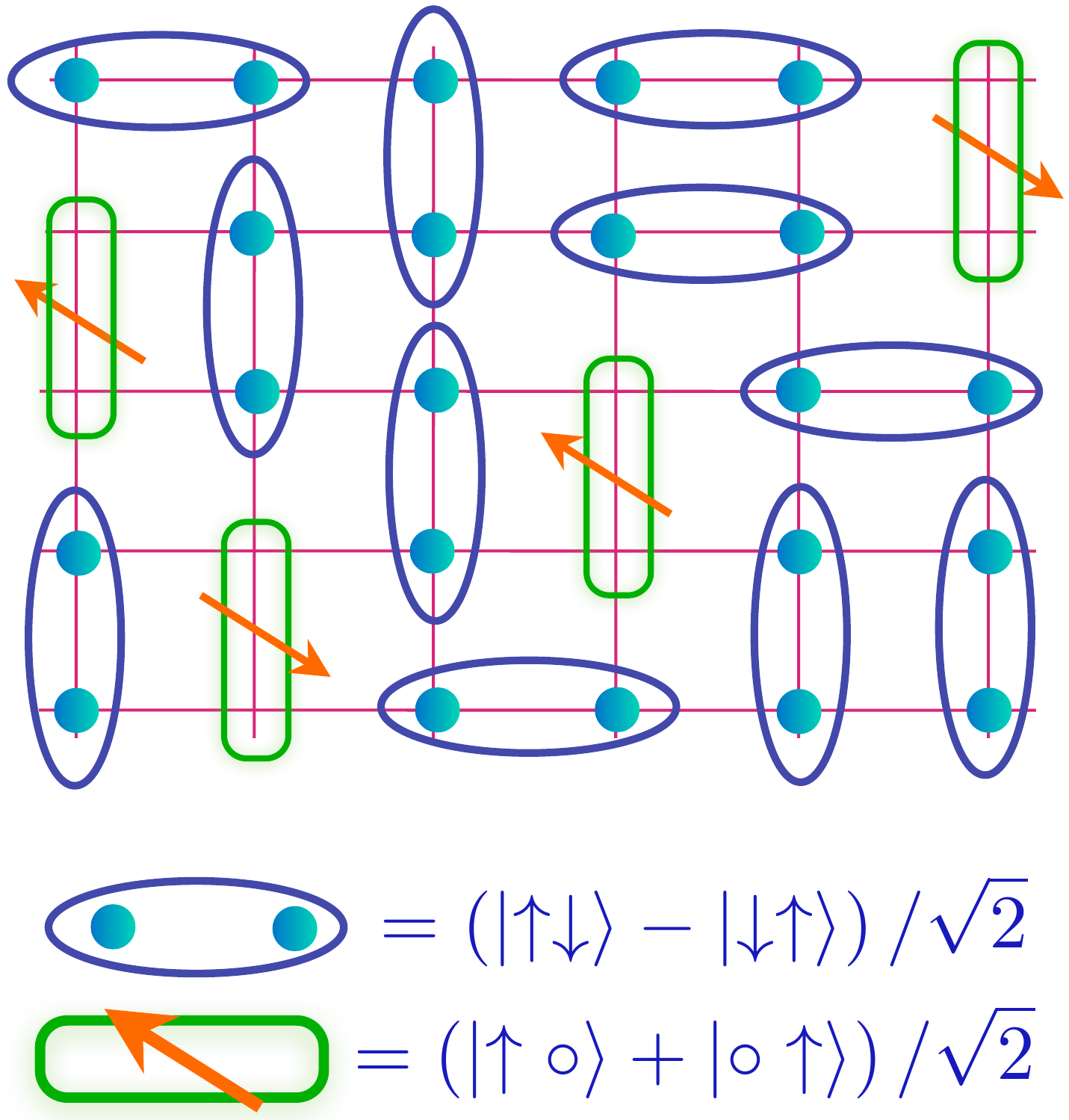}
	\caption{A state in the Hilbert state of the dimer model. The blue dimers are bosons representing a spin-singlet
	pair of electrons. The green dimers are spin-1/2 fermions representing an electron in a bonding orbital between
	a pair of sites.}
	\label{fig:dimer}
\end{figure}

The quantum dimer model of Ref.~[\onlinecite{Punk15}] 
has bosonic dimers and spin-1/2 fermionic dimers, which close pack a square lattice with an even number of sites (see Fig.~\ref{fig:dimer}).
The bosonic sector of this model is identical to that of the original study of Rokhsar and Kivelson (RK)~\cite{DRSK88}, with a potential and resonating term for dimers within a plaquette. 
In addition, fermionic dimers may move via hopping terms whose form will be specified below. 
Interaction between the fermionic dimers can, in principle, be present but is not expected to be important 
when the density of fermions is low; we will not include fermion-fermion interactions here.

Now we state the Hamiltonian for this model. 
Let us first define the operators creating (annihilating) bosonic and fermionic dimers as $D^{\dagger}_{ix}$ ($D_{ix}$) and $F^{\dagger}_{ix\alpha}$ ($F_{ix\alpha}$), respectively. 
The extra indices $i$ and $x$ ($y$) indicate the created or annihilated dimer resides on the link between $i = (i_x , i_y)$ and $i + {\hat x}({\hat y})$, where
$\hat x = (1, 0)$ and $\hat y = (0,1)$ are unit vectors and $\alpha = \uparrow , \downarrow$ is the spin index.
Note that we set the lattice spacing to 1. 
In the language of the $t$-$J$ model, $D$ and $F$ operators have the following correspondence to the electron creation and annihilation operators $c^{\dagger}$, $c$:
\begin{align}
	D^{\dagger}_{i\eta} &\sim \frac{(-1)^i}{\sqrt{2}} 
	\left( c^{\dagger}_{i \uparrow} c^{\dagger}_{i + \hat \eta , \downarrow} 
	+ c^{\dagger}_{i \downarrow} c^{\dagger}_{i + \hat \eta , \uparrow} \right), \nn
	F^{\dagger}_{i\eta\alpha} &\sim \frac{(-1)^i}{\sqrt{2}} 
	\left( c^{\dagger}_{i \alpha}	+ c^{\dagger}_{i + \hat \eta , \alpha}  \right).
\end{align}
Here, $(-1)^i$ is due to a gauge choice which we follow from Ref.~[\onlinecite{DRSK88}]. 
We can observe that the quantum numbers of states $D^{\dagger}_i \vac$ and $F^{\dagger}_{i\alpha} \vac$ are the same as $c^{\dagger}_{i \uparrow}c^{\dagger}_{i \downarrow} \vac$ and $c^{\dagger}_{i \alpha} \vac$, setting aside the fact that the degrees of freedom of the former live between two sites ($i$ and $i + \hat \eta$) and the latter reside on each site. 
This fact will be useful in our DMRG setup. 
We can now write the Hamiltonian for the model in terms of dimer creation and annihilation operators~\cite{Punk15},
\begin{align}
	H &= H_{\textrm {RK}} + H_1 ,\nn
	H_{\textrm {RK}}
	&= \sum_i \left[ - J \,D^{\dagger}_{ix} D^{\dagger}_{i+{\hat y}, x} D_{iy} D_{i+{\hat x}, y} 
	+ {\textrm {1 term}} \right. \nn
	& \mathrel{\phantom{= \sum_i [}} \left. +  V \,D^{\dagger}_{ix} D^{\dagger}_{i+{\hat y}, x} D_{ix} D_{i+{\hat y}, x} 
	+ {\textrm {1 term}} \right] ,\nn
	H_1 &= \sum_{i,\alpha} \left[ - t_1 \, D^{\dagger}_{ix} F^{\dagger}_{i+{\hat y}, x \alpha} 
										F_{ix\alpha} D_{i+{\hat y}, x} + {\textrm {3 terms}} \right. \nn
	& \mathrel{\phantom{= \sum_i [}} - t_2  \,  D^{\dagger}_{i + {\hat x},y} F^{\dagger}_{iy \alpha}  
									 			F_{ix\alpha} D_{i+{\hat y}, x} + {\textrm {7 terms}} \nn
	& \mathrel{\phantom{= \sum_i [}} - t_3 \,D^{\dagger}_{i + {\hat x} + {\hat y},x} F^{\dagger}_{iy \alpha}  
									 		F_{i + {\hat x} + {\hat y}, x\alpha} D_{iy} + {\textrm {7 terms}} \nn
	& \mathrel{\phantom{= \sum_i [}} \left .- t_3  \,  D^{\dagger}_{i + 2{\hat y},x} F^{\dagger}_{iy \alpha}  
									 			F_{i + 2{\hat y}, x\alpha} D_{iy} + {\textrm {7 terms}} \right].
	\label{dimer_model}
\end{align} 
The terms we have not explicitly written down are connected to the previous term through a symmetry transformation of the square lattice. 
$H_{\textrm{RK}}$ is the pure bosonic sector mentioned above; $J$ is the coupling for the resonant term, and $V$ is the coupling for the potential term. 
$H_1$ contains the hopping terms of the fermionic dimers; $t_1$, $t_2$, and $t_3$ correspond to three distinct types of hoppings. 

In the absence of the fermions, the undoped dimer model has an exactly solvable point (the RK point) at $V=J$,
with a spin-liquid ground state given by the equal superposition state of all allowed $D_i$ dimer 
configurations \cite{DRSK88}. But away from this point, bipartite dimer models are described by a
dual compact U(1) gauge theory \cite{EFSK90} and have been argued to have only confining valence-bond solid ground 
states \cite{NRSS90,AVLBTS04,EFSLS04}.  Adding a finite density of fermions is not expected to change the basic
structure of the bosonic dimer model, apart from modifying their short-distance effective action \cite{PCAS16}.
However, the confinement length scale is
large near the RK point and at small fermion density, and a spin-liquid U(1)-FL* state should be effectively realized when the confinement
scale is larger than the system size. We study such a regime in the present paper, and for our system sizes, our results are consistent
with deconfinement. We expect such results to apply to the higher-temperature pseudogap in the cuprates, where a thermal scale
cuts off the crossover to confinement. At lower temperature, there will be a crossover to confinement and translational symmetry
breaking, as has been discussed in recent work \cite{PCAS16}. Alternatively, deconfinement can survive in quantum dimer models without
a sublattice structure \cite{SSkagome} with $\mathbb{Z}_2$ topological order, but we will not consider such dimer models here.

In our DMRG calculation, we consider a lattice with geometry of a finite cylinder. 
The circumference of the cylinder consists of four lattice sites, and the length of the cylinder is up to 32 sites. 
In the second part of this paper, we also compute entanglement entropies in a $64 \times 2$ cylinder to observe one-dimensional effects.
We repeat our calculation in different fermionic dimer densities, from one to eight fermionic dimers.
Note that in our lattice configuration, eight fermionic dimers correspond to $1/16$ doping in the typical cuprate phase diagram. 
We use two sets of parameters for the couplings in Eq.~(\ref{dimer_model}). 
One is the parameters which are relevant to the physical model for the cuprates in the pseudogap regime: $t_1 = -1.05 J$, $t_2 = 1.95 J$, and $t_3 = -0.6 J$, near the RK point ($V = 0.9J$). 
The other is the parameter which we choose for comparison: $t_1 = t_2 = t_3 = J$, also with $V = 0.9J$. 
The single-fermion study in Ref.~[\onlinecite{Punk15}] suggests the different hopping parameters change the dispersion of the fermionic dimer: the Fermi surface consists of four hole pockets near $(\pm \pi/2 , \, \pm \pi/2)$ in the former parameter regime and a single Fermi surface centered at $(0,0)$ in the latter. 
We will confirm this behavior in our DMRG calculation below, while studying a multiple-fermion system.

The fact that we are interested in observing the hole pockets is closely related to the reason we chose the circumference as a four-lattice site for the first part of our calculation. 
The center of the four hole pockets is at $\vec{k} \sim (\pm \pi/2 , \, \pm \pi/2)$, and the minimum number of sites in the $y$ direction needed to get information about $k_y = \pm \pi/2$ is four.
This is why we could not choose two sites in the circumference when our focus is on the fermion dispersion. 
Later, when we concentrate on the one-dimensional scaling properties of the entanglement entropies, we study the case of a cylinder of two sites along the circumference, which allows us to calculate much longer system. 

Now we comment on the topological sectors of the Hamiltonian. 
The Hilbert space we are considering consists of closely packed configurations of dimers. 
For each configuration, we can define an integer quantity 
\beq
w_x = \sum_{i_x =1}^{N_x} (-1)^{i_x} (D^{\dagger}_{(i_x, i_y)y} D_{(i_x, i_y)y} + F^{\dagger}_{(i_x, i_y)y\alpha} F_{(i_x,i_y)y\alpha} )
\eeq
for any $1 \leq i_y \leq N_y$, where $N_x$ ($N_y$) is the number of sites in the $x$ ($y$) direction and the spin $\alpha$ is implicitly summed. 
One can observe that every term in Eq.~(\ref{dimer_model}) preserves this quantity, so the possible configurations of the dimers spanning the Hilbert space can be divided into different sectors with different values of $w_x$.  
The integer $w_x$ is the topological winding number associated with loops in the transition graph that circles the $x$ axis, similar to that of the RK model on a torus~\cite{DRSK88}.
The integer $w_y$ can be defined in an analogous manner, but it is not meaningful since we do not have periodic boundary conditions in the $x$ direction.
In principle, we would like to restrict ourselves to the zero-winding-number sector, $w_x = 0$.
Right at the RK point, $J=V$, each topological sector has a unique ground state with zero energy, which is an equal 
superposition of all configurations \cite{DRSK88}.
Since the number of configurations is largest at $w_x = 0$, sectors with a large absolute value of $w_x$ have lower entanglement and will be preferred by DMRG.
One way of imposing the $w_x = 0$ condition is to add a potential term proportional to $w_x^2$ to the Hamiltonian at the price of more computation. 
Another method is to tune $V$ to be slightly smaller than $J$ and penalize states moving away from $w_x = 0$. 
In this case, the amount of computation is similar to that without the constraint because we do not add any additional terms to the Hamiltonian. However, we have to choose an optimal value for the $V/J$. 
We adopt the latter method since moving away from the fine-tuned RK point is beneficial for us, and in all of our calculations, we have used $V = 0.9 J$. The confinement length scales are large enough at this coupling that we do not observe
valence-bond solid order, even in the undoped case.

Each bond in the square lattice can have four states: 
occupied by a bosonic dimer, occupied by a spin-up or -down fermionic dimer, or empty. 
As mentioned previously, one useful observation is that the quantum numbers of the bonds are the same as the quantum numbers of the sites in a spin-half fermion model, with the bosonic dimer occupied state corresponding to the filled (spin-up and -down) state. 
Therefore we can map our dimer model to a fermionic Hubbard model on the links, with dimer constraints. 
The dimer constraint is to ensure each site is part of only one dimer, and this is achieved in our DMRG as an additional potential term. 
We use a potential of $\sim 20J$ to penalize overlapping dimers. 

All DMRG calculations in this paper were performed with the ITensor library \footnote{ITensor library, version 1.2.0, {\tt http://itensor.org/}}.
We kept up to $\sim 600$ states to keep the truncation error per step to $\sim 10^{-8}$. 

\section{Density Modulation}

Now we show the results of the DMRG calculations where we can observe the change in dispersions with different hopping parameters, especially the existence of a dispersion with Fermi pockets centered near $\vec k = (\pm \pi/2 , \pm \pi/2)$. 
Extracting momentum information from DMRG is not trivial since it is a real-space calculation. (For few fermions, one can use techniques used in Ref.~[\onlinecite{SWDSSK15}] to achieve this. Also, there are more recent schemes for DMRG in mixed real and momentum space proposed in Ref.~[\onlinecite{Motruk16}].)
However, we may observe Friedel oscillations from the open boundaries of our system, and these will reveal information of the fermionic dimer's momentum in the cylinder direction. 

\begin{figure}
\centering
	\vspace*{1mm}
	\includegraphics*[width=87mm,angle=0]{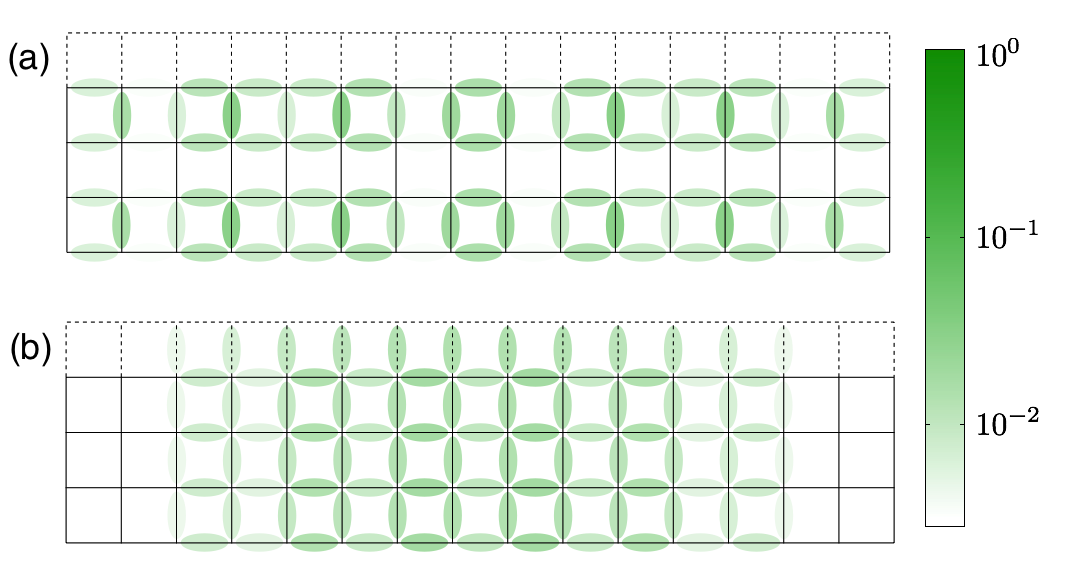}
	\caption{The log-scale density of fermionic dimers on a $16 \times 4$ lattice. The configuration consists of a single fermionic dimer and 31 bosonic dimers. The dashed line indicates the periodic boundary condition in the $y$ direction; the top dashed line is identified with the bottom solid line. The hopping parameters used are (a) $t_1 = -1.05 J$, $t_2 = 1.95 J$, $t_3 = -0.6 J$ and (b) $t_1 = t_2 = t_3 = J$. In (a), one can observe the density oscillation with a period of roughly two lattice sites, which corresponds to crystal momenta of $\pi/2$. Note that the Hilbert space is a closely packed dimer configuration, and sites without fermionic dimers are occupied by bosonic dimers.  }
	\label{fig:friedel}
\end{figure}

First, we check whether the Friedel oscillation observed in the case of a single fermionic dimer is consistent with Ref.~[\onlinecite{Punk15}].
Figure~\ref{fig:friedel} is the density profile of the fermionic dimers when a single fermionic dimer is present among bosonic dimers on a $16 \times 4$ lattice for the two parameter sets we use. 
From Fig.~\ref{fig:friedel}(a), which is the parameter set expected to have hole pockets, we can observe an oscillation of the profile starting from the open boundary to the $x$ direction.
This is especially clear when looking at the vertical dimers. 
The period of the oscillation is roughly two lattice sites.
Since the Friedel oscillation has a wave vector of $2 k_F$, this indicates that the fermionic dimer in the ground state has a crystal momentum of $k_x \sim \pi/2$. 
This fact is consistent with the exact diagonalization study in Ref.~[\onlinecite{Punk15}], which found the energy minima of the single-fermion spectrum to be near $\vec k = (\pi/2 , \pi/2)$. 
On the other hand, Fig.~\ref{fig:friedel}(b) does not show any prominent oscillation near the boundary. 
This calculation has been done with the parameters which are expected to have a single band with the dispersion minima at $\vec k = (0 , 0)$, so the absence of Friedel oscillation is expected. 
We have performed the same calculation for a single fermionic dimer while increasing the $x$ direction of the lattice size, up to a $32 \times 4$ lattice, and have observed the same behavior, in both cases with dispersion minima at $k_x = \pi/2$ [Fig.~\ref{fig:friedel}(a)] and $k_x = 0$ [Fig.~\ref{fig:friedel}(b)].
A more quantitative analysis for the $32 \times 4$ lattice by Fourier transform will follow below, together with the higher-density calculation.

Note that Fig.~\ref{fig:friedel}(a) seems to break the translation symmetry in the $y$ direction. 
However, this is just a spontaneous symmetry breaking between the two degenerate ground states, one being Fig.~\ref{fig:friedel}(a) and the other being Fig.~\ref{fig:friedel}(a) translated by one lattice site in the $y$ direction.
One can check this by computing the ground state several times and obtaining both states or by adding a small perturbation acting as a chemical potential on vertical bonds with a particular $y$ coordinate and observing the absence of the symmetry.
The fact that Fig.~\ref{fig:friedel}(b) does not break the symmetry is also in accordance with our claim. 
Since the state of Fig.~\ref{fig:friedel}(b) has 
only one Fermi surface centered at $\vec{k} = (0 , \, 0)$, there is no degeneracy in the ground state. 

\begin{figure}
\centering
	\vspace*{1mm}
	\includegraphics*[width=87mm,angle=0]{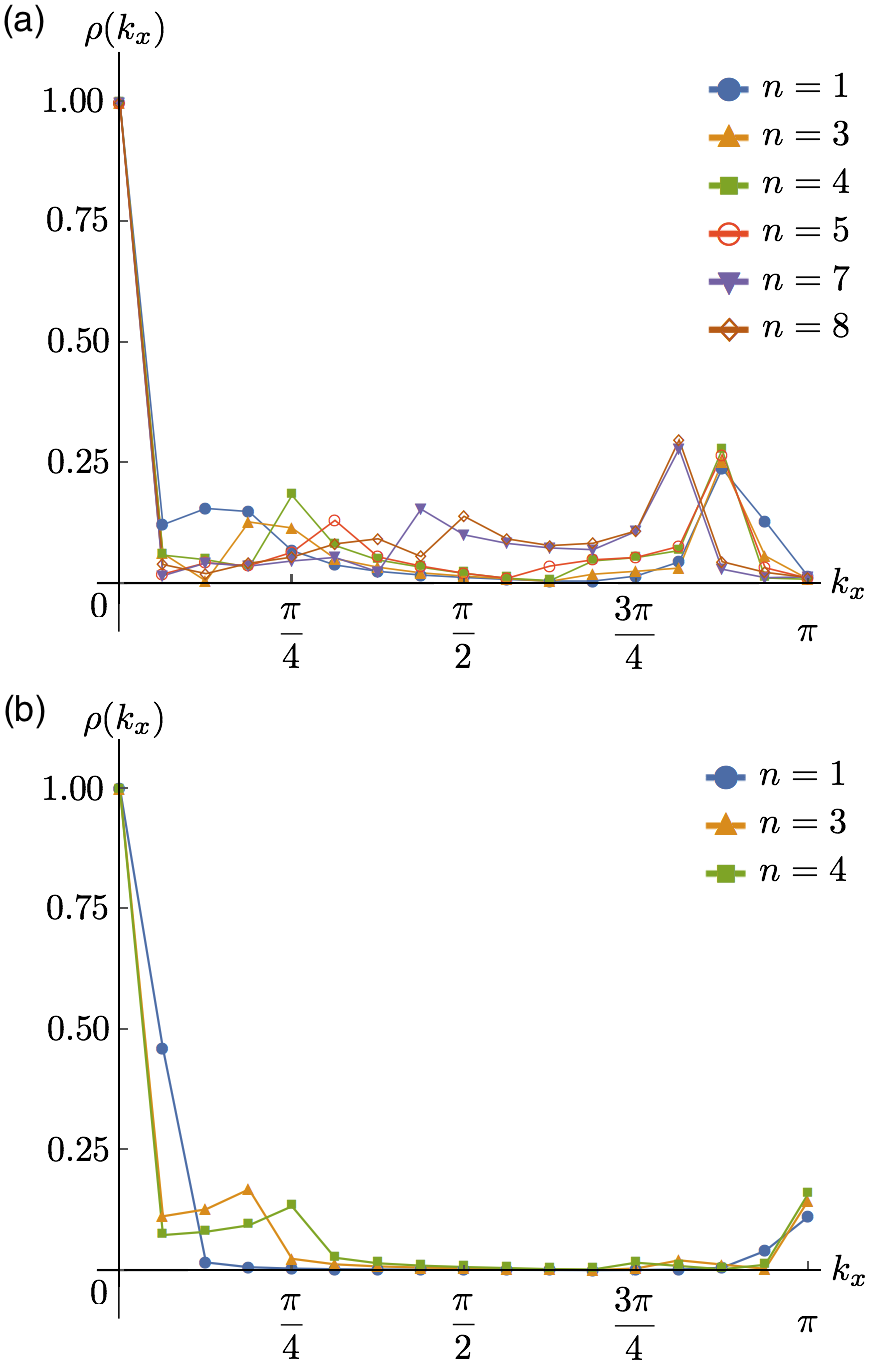}
	\caption{Fourier transform of the density of fermionic dimers with various total densities. $n$ denotes the number of fermionic dimers in the system. Note that in our $32 \times 4$ lattice, $n=8$ corresponds to $1/16$ doping. The hopping parameters used are (a) $t_1 = -1.05 J$, $t_2 = 1.95 J$, $t_3 = -0.6 J$ and (b) $t_1 = t_2 = t_3 = J$. In (a), there are peaks at $13 \pi/16$ and $7 \pi/8$, which indicates the fermionic dimers with $k_x = 13 \pi /32$ and $7 \pi /16$ are at the Fermi level, which is a feature missing in (b). The central peak at zero is due to the total density and is normalized to 1. }
	\label{fig:fourier}
\end{figure}

We would like to study the Friedel oscillation more quantitatively and verify whether this feature survives when we increase the number of fermionic dimers $n$. 
We keep the lattice size as $32 \times 4$ and increase $n$ up to 8, which corresponds to $1/16$ doping. 
Since the ``defects'' of the system are the open boundaries, the Friedel oscillation is in the cylinder direction.
From the density profile $\rho(x, y)$, we define $\rho_x (x) = \sum_y \rho(x,y)$ and perform Fourier transformation.
The result is shown in Fig.~\ref{fig:fourier}. 
Note that we have normalized the data by $1/n$, and the magnitude-1 peak at $k_x = 0$ indicates the total density is $n$. 
Other than the $k_x = 0$ peak, we can observe that there is a peak at $k_x = 7 \pi /8$ in $n = 1, 3, 4, 5$ in Fig.~\ref{fig:fourier}(a), where the parameter set used is the same as in Fig.~\ref{fig:friedel}(a). 
This peak is due to the Friedel oscillation and indicates that $k_x = 7 \pi / 16$ at the Fermi level. 
Reference~[\onlinecite{Punk15}] showed the energy minimum is at $\vec{k} = (q, q)$ for $q$ slightly less than $\pi/2$, and this is in good agreement with our result.
Based on experiments and previous works, one can argue the energy minimum should be either at the origin [$\vec{k} = (0, 0)$] or along the diagonal [$\vec{k} = (q, q)$].
Therefore, we can conclude the dispersion of the dimer model in our cylinder will have a minimum at $\vec{k} = (7 \pi / 16, \pi/2)$, and in the large system limit this will converge to a diagonal point $\vec{k} = (q, q)$ with $7 \pi / 16 \leq q \leq \pi/2$. 
Moreover, from $n = 7, 8$ data in Fig.~\ref{fig:fourier}(a), we can observe the expansion of the Fermi surface as we increase the fermionic dimer density. The new peak at $k_x = 13 \pi /16$ indicates that the Friedel oscillation is now from the new Fermi level at $k_x = 13 \pi /32$. 

\begin{figure}
\centering
	\vspace*{1mm}
	\includegraphics*[width=87mm,angle=0]{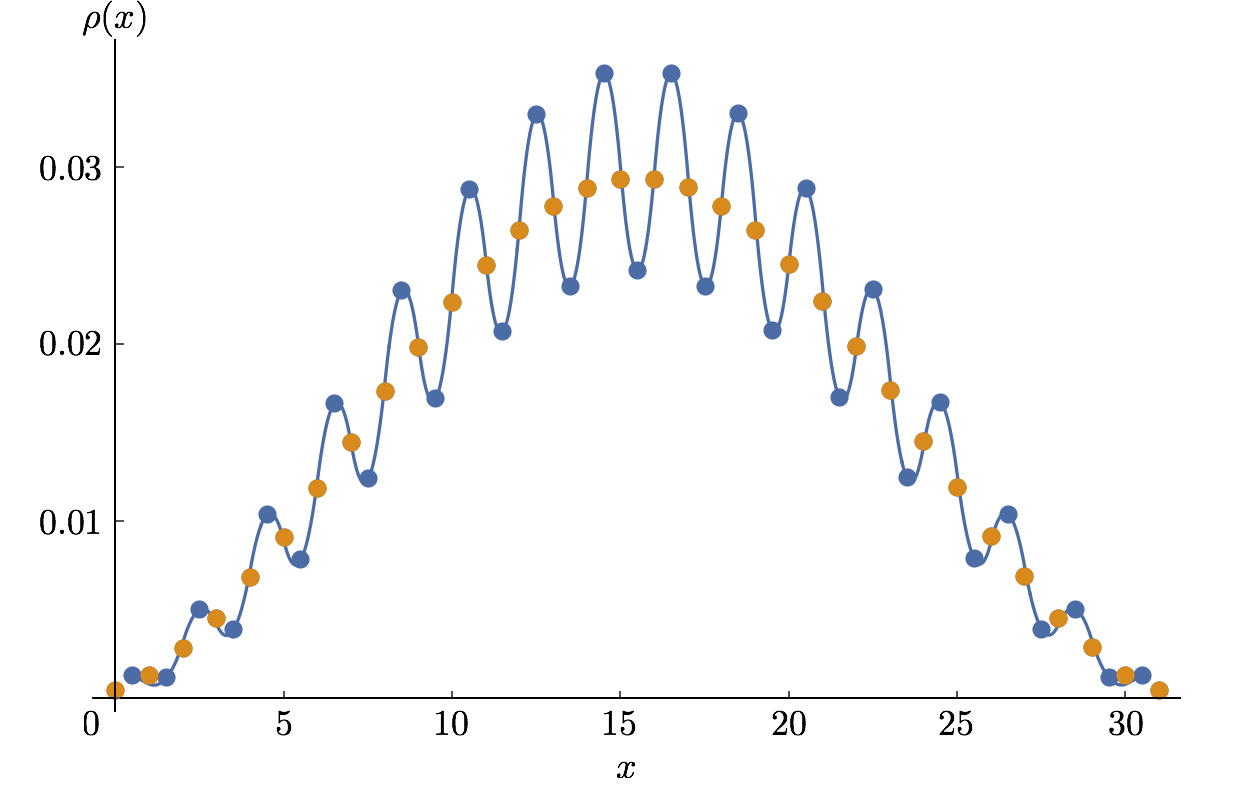}
	\caption{One-dimensional plot for fermionic dimer density [$\rho (x) = \sum_y \rho(x,y)$] as a function of distance in the $x$ direction when $n=1$. The $x$-axis unit is the lattice constant. Vertical bonds have integer $x$ and are colored yellow; horizontal bonds have half-integer $x$ and are colored blue. Notice the modulation is only present in the horizontal dimers. The system is a $32 \times 4$ cylinder with parameters $t_1 = t_2 = t_3 = J$.}
	\label{fig:den_pm2}
\end{figure}

For Fig.~\ref{fig:fourier}(b), which uses the same parameter set as Fig.~\ref{fig:friedel}(b), there is no peak at $k_x = 7 \pi /8$ for any $n$. 
This is in accordance with our expectation that the state has a dispersion minimum at $\vec{k} = (0, \, 0)$ and also with the qualitative result we have seen in Fig.~\ref{fig:friedel}(b).
There is a signal at $k_x = \pi$; however, the origin of this signal is not the Friedel oscillation. 
Figure~\ref{fig:den_pm2} shows the density of the fermionic dimer as a function of $x$. 
The plotted one-dimensional density $\rho (x) = \sum_y \rho(x,y)$ is the Fourier transform of $\rho (k_x)$, which is the quantity plotted in Fig.~\ref{fig:fourier}.
The $x$ axis of the plot is the position in units of lattice constant.
Integer values are for the vertical bonds, and half integers are for the horizontal bonds. 
Looking at only the vertical bonds does not show any modulation in the density, and the pattern looks very much like a particle in a box. 
On the other hand, the horizontal bond shows some modulation with two lattice sites. 
This density modulation clearly has a wave vector of $\pi$ and is the reason for the signal at $k_x = \pi$ in Fig.~\ref{fig:fourier}(b).
Although the precise reason for this oscillation is unclear, we 
can clearly see that this modulation is present throughout the bulk and the signal at $k_x = \pi$ does not indicate Friedel oscillation from $k_F = \pi/2$: it appears to be simply a lattice commensuration effect. 

Note that we have not included the data obtained for $n$ with $\textrm {mod} \,(n, 4) = 2$. 
The calculations with such $n$ had a stronger tendency towards the $w_x \neq 0$ topological sector, and we had to decrease $V$ further to keep the state in the $w_x = 0$ sector. (For $n=2$, we needed $V < 0.8$.)
This seems to be an artifact of our system, which is effectively one-dimensional and can have only four values of $k_y$.

\section{Entanglement entropy}

We present the result for the computation of R\'{e}nyi entropy to gain more information about the ground state of the dimer model.
First, recall the definition of the $\alpha$th R\'{e}nyi entanglement entropy:
\begin{align}
	S_\alpha = \frac{1}{1 - \alpha} {\textrm {ln}} \left[ {\textrm{Tr}}\, \rho^\alpha_A \right].
\end{align}
Here, $\rho_A$ is the reduced density matrix of partition $A$, i.e., $\rho_A = \textrm {Tr}_B\, \rho$, where $A \cup B$ is the total system.
Note that the $\alpha$th R\'{e}nyi entropy becomes the von Neumann entropy in the $\alpha \rightarrow 1$ limit. 

In a one-dimensional gapless system, conformal field theory (CFT) has a result for the scaling of the R\'{e}nyi entropy~\cite{Cardy04, Calabrese10}:
\begin{align}
	S_\alpha = \frac{c}{12} \left( 1 + \frac{1}{\alpha} \right) \textrm{ln} 
	\left( \frac{2L}{\pi} \, \textrm{sin} \frac{\pi l}{L} \right) + g + c'_\alpha. 
	\label{eq:cft_ent}
\end{align}
This is the case for a finite system of length $L$ with open boundary conditions, divided into two pieces, where the length of one piece is $l$. 
$g$ is the boundary entropy~\cite{Affleck91}, and $c'_\alpha$ is a nonuniversal constant. 
Considering our system to be quasi-one-dimensional, we can extract the central charge $c$ of the system from this equation. 
For the entanglement entropy calculation, we consider both $32 \times 4$ and $64 \times 2$ cylinders to see any scaling behavior as the system approaches one dimension. 
We partition the system into two cylinders, each of length $l$ and $L - l$ in the context of Eq.~(\ref{eq:cft_ent}), by a single cut in the cylinder direction.
Moreover, now we concentrate on the parameters which give a single Fermi surface, $t_1 = t_2 = t_3 = J$.
The results for the other parameters are expected to be four copies of the presented results in the thermodynamic limit. 
However, the density modulations resulting from the open boundaries prevent the entanglement entropy data from having a nice one-dimensional scaling form for $L = 32$.

\begin{figure}
\centering
	\vspace*{1mm}
	\includegraphics*[width=87mm,angle=0]{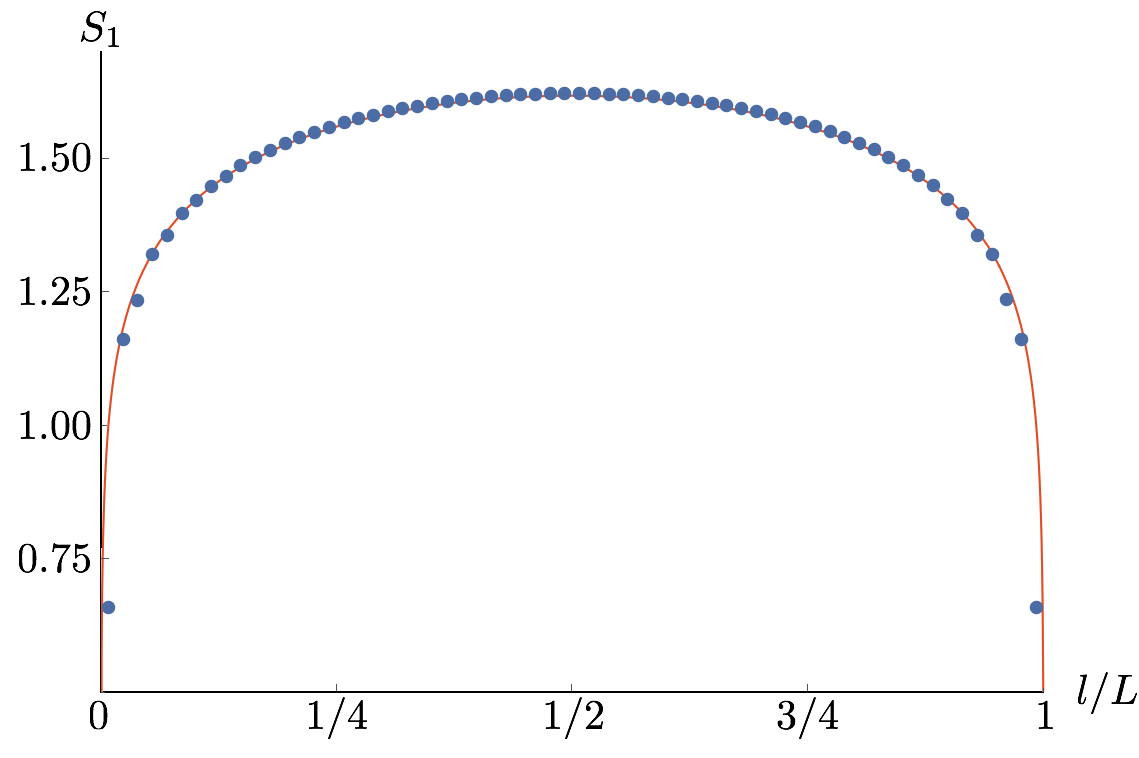}
	\caption{Von Neumann entropy of the pure RK model, calculated on a $64 \times 2$ cylinder. The red solid line is not the interpolation of the data points but the exact CFT result of Eq.~(\ref{eq:cft_ent}) with $c=1$ and $g + c'_\alpha = 1$. }
	\label{fig:RKentropy}
\end{figure}
First, we show the comparison of the von Neumann entropy of the pure RK model, with $n=0$ fermions, with the CFT result
(\ref{eq:cft_ent}) in Fig.~\ref{fig:RKentropy}. An excellent fit is found for $c=1$. The fermion-free dimer model is dual to a sine-Gordon 
model \cite{EFSK90,NRSS90},
and in 1+1 dimensions this has a gapless phase described by a massless relativistic boson with $c=1$. The results in
Fig.~\ref{fig:RKentropy} are in accordance with this expectation.

\begin{figure}
\centering
	\vspace*{1mm}
	\includegraphics*[width=87mm,angle=0]{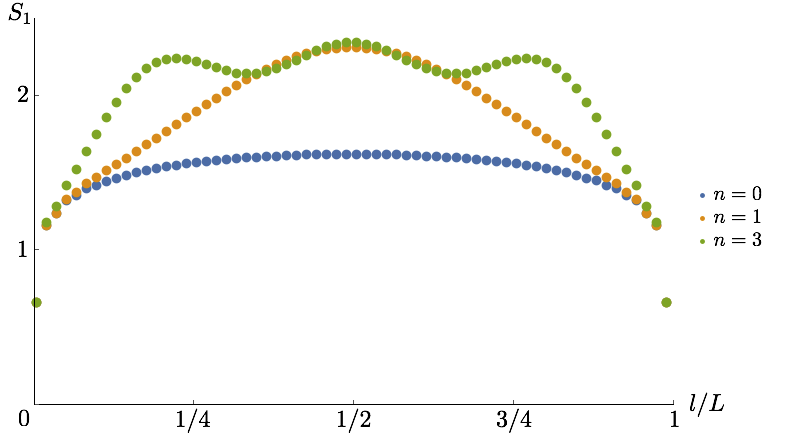}
	\caption{Von Neumann entropy of the FL* phase with different fermionic dimer densities. The system is a $64 \times 2$ cylinder with parameters $t_1 = t_2 = t_3 = J$. $L$ is the length of the system, and $l$ is the length of the subsystem. In the case of $n = 0$, which is a pure bosonic dimer model near the RK point, we get a nice fit to Eq.~(\ref{eq:cft_ent}) with central charge 1.}
	\label{fig:entropy}
\end{figure}

\begin{figure}
\centering
	\vspace*{1mm}
	\includegraphics*[width=87mm,angle=0]{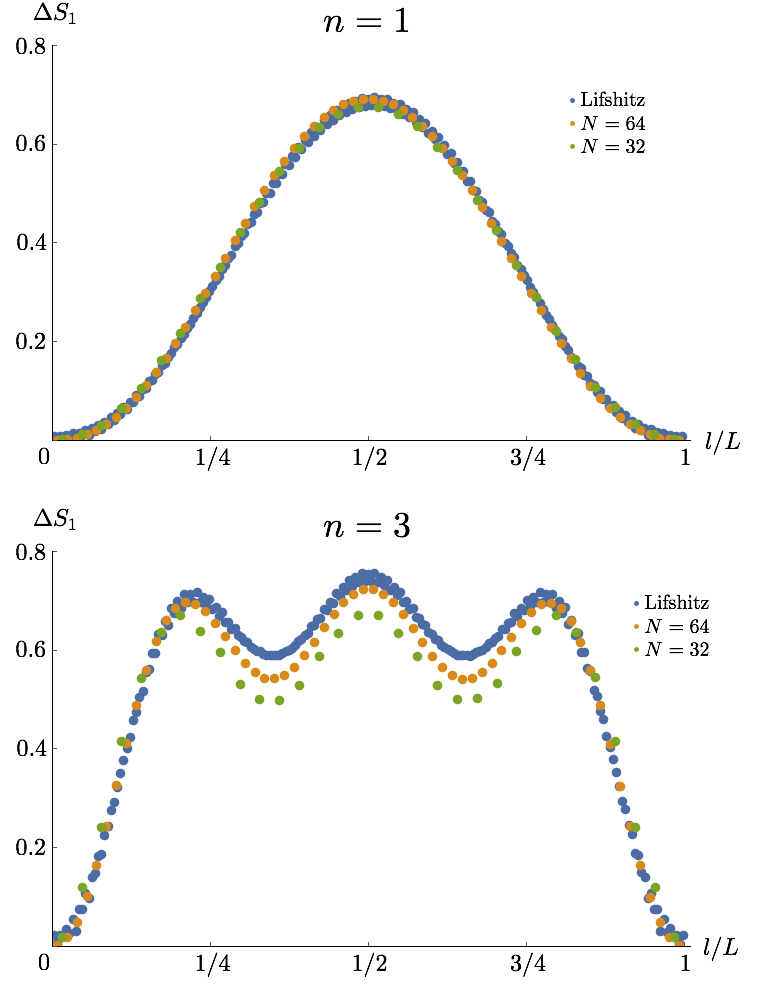}
	\caption{Fermion contribution to the von Neumann entropy of the dimer model and the Lifshitz transition. $\Delta S_1$ equals $S_1 (n = 1) - S_1 ( n = 0)$ on the left and $S_1 (n = 3) - S_1 ( n = 0)$ on the right, where $n$ is the number of fermionic dimers. 
	The system is a $64 \times 2$ and $32 \times 4$ cylinder with parameters $t_1 = t_2 = t_3 = J$ for the dimer model and free fermions on a 200-site chain for the Lifshitz transition. The $n=1$ and $n=3$ cases for Lifshitz transition correspond to the number of occupied states in the new band.}
	\label{fig:entropy2}
\end{figure}

Turning to the case with fermions, the von Neumann entropy when the fermion hopping parameters are set to $t_1 = t_2 = t_3 = J$ is shown in Fig.~\ref{fig:entropy}.
The results are for a $64 \times 2$ lattice. 
Data for the $32 \times 4$ lattice are not shown but are very similar to the presented data ($32 \times 4$ results are included in Fig.~\ref{fig:entropy2}).
We also include the data from Fig.~\ref{fig:RKentropy} for the case without any fermions. It is clear that these is an additional
contribution from the presence of the fermions, but it cannot be accounted for by changing the central charge of the CFT. 
Fermions at nonzero density in an infinite system should form a Fermi surface, and in the quasi-one-dimensional geometry,
each Fermi point should yield an additional contribution of $c=1/2$ of a chiral fermion. It is clear that the 
data in Fig.~\ref{fig:entropy}
are not of this form.  

Instead, we found that an excellent understanding of Fig.~\ref{fig:entropy} is obtained by thinking about 
the limit of a very low density of fermions at the bottom of a quadratically dispersing band. 
This is the case of a ``Lifshitz'' transition in one dimension, when the chemical potential crosses the bottom
of a band. 
The authors of Ref.~[\onlinecite{lifshitz}] studied the entanglement entropy near such a Lifshitz transition. 
In their Fig.~11, they present the entanglement entropy of a half-filled free-fermion system with 200 sites, as the next-nearest hopping $t$ is tuned to go across the Lifshitz transition. 
[The Hamiltonian used is $H = -\sum_i ( c^\dagger_i c_{i + 1} + t c^\dagger_i c_{i + 2}) + \textrm{H.c.}$]
Different graphs are labeled by different values of $t$, but basically, what is changing is the number of occupied states above the Lifshitz transition. 
For example, when $t = 0.5$, only the large Fermi surface is occupied; when $t = 0.51$, the system has just gone through a Lifshitz transition, and one state is occupied from the new band; when $t = 0.52$, two states are occupied above the Lifshitz transition. 
The number of modulations in the entanglement entropy exactly matches the number of states filled above the Lifshitz transition. 

We reproduced the data of Fig.~11 in Ref.~[\onlinecite{lifshitz}] to compare with the behavior of the entanglement entropy of our own system of fermionic and bosonic dimers.
Figure~\ref{fig:entropy2} shows the fermionic contribution $\Delta S_1$ of the entanglement entropy. 
This is obtained by subtracting the entanglement entropy of $n = 0$, which was shown in Fig.~\ref{fig:RKentropy} to
be due to a $c = 1$ boson field.
To compare with the case of Lifshitz transition, we also subtract the entanglement entropy of the system with only one large band occupied, which is $t = 0.50$ in the specific model, from the system with one (three) state(s) occupied in the new band, corresponding to $t = 0.51$ (0.53); in this case, the gapless fermions from the occupied large band contribute as a $c = 1$ field.
As seen in Fig.~\ref{fig:entropy2}, $\Delta S_1$ for $n = 1$ is nearly identical to the corresponding entanglement
entropy for the Lifshitz transition for free fermions with two different lattice sizes. 
For $n = 3$, the value of $\Delta S_1$ decreases slightly as the length of the system decreases, but the qualitative features remain the same.
Note that in these data, only the total length of the system was scaled to unity. 

The above results provide strong evidence that the dimer model can be viewed as two approximately independent systems:
a background $c=1$ boson corresponding to the resonance between the dimers (both blue and green \cite{PCAS16}) 
and a dilute gas (of density $p$) of free 
fermions. These are precisely the characteristics of the FL* state, which has an emergent gauge field (represented here by
the $c=1$ boson) and a Fermi surface of electronlike quasiparticles. 

\section{Outlook}

The combination of our results on the density distribution and the entanglement entropy confirms the expected 
appearance
of a FL* state in the dimer model of Ref.~[\onlinecite{Punk15}]. By general arguments \cite{TSMVSS04,APAV04}, 
the violation of the Luttinger theorem for a Fermi liquid requires that the emergent gauge fields appear in the spectrum
of the theory. Our results on the entanglement entropy in a quasi-one-dimensional geometry are in accordance
with this requirement, showing a background $c=1$ boson that is expected from the gauge theory of the dimer 
model \cite{EFSK90,NRSS90}; the boson represents the modes associated with the ``resonance'' between the dimers
around a plaquette. Above this gauge field background, we obtained evidence of a gas of nearly free fermions of density $p$,
both in the density modulations and in the entanglement entropy: in particular, the fermionic contribution to the entanglement entropy closely matched that of a dilute gas of free fermions near the bottom of a quadratically dispersing band.

We can extend our calculation to wider systems, which can reveal the properties of the model in two dimensions and also provide us with a better momentum resolution in the $k_y$ direction.
With the off-diagonal measurement between different particle number states on these wider cylinders~\cite{SWDSSK15}, we can measure Fermi surface and quasiparticle residue with good resolution.
Another possible approach is to use the recent proposal of DMRG in mixed real and momentum space~\cite{Motruk16}.
With this method we will be able to determine $k_y$ more directly, which can be complimentary to the off-diagonal measurements.
Moreover, Ref.~[\onlinecite{Motruk16}] claims the computation time is reduced by more than an order of magnitude, so this may allow us to include even more sites in the $y$ direction (in this case, equivalently, more $k_y$ points), resulting in better $k_y$ resolution.
Moreover, dynamical information about the model can be calculated from a time-dependent DMRG calculation~\cite{Vidal2003,SWAF04,ADCKUSGV04}.

\subsection*{Acknowledgments}

We thank A. Allais, M. Punk, B. Swingle, and W. Witczak-Krempa for useful discussions. 
J.L. is particularly grateful to M. Stoudenmire for introducing and teaching DMRG and the ITensor library. 
The research was supported by the NSF under Grants No. DMR-1360789 (J.L., S.S.) and No. DMR-1505406 (S.R.W.) and MURI Grant No. W911NF-14-1-0003 from ARO (J.L., S.S.).
Research at the Perimeter Institute is supported by the government of Canada through Industry Canada and by the province of Ontario through the Ministry of Research and Innovation. 

\bibliography{cuprates}

\end{document}